\documentclass[10pt,aps,prd,amsmath,amssymb,twoside,showpacs,nofootinbib,preprintnumbers]{revtex4}
\usepackage[utf8]{inputenc}
\usepackage[T1]{fontenc}
\usepackage{lmodern}

\newcommand{\defeq}{:=}
\newcommand{\im}{\mathrm{i}}  

\newcommand{\R}{\mathbb{R}}
\newcommand{\C}{\mathbb{C}}
\newcommand{\xd}{\mathrm{d}}
\newcommand{\xD}{\mathcal{D}} 

\newcommand{\cH}{\mathcal{H}}
\newcommand{\tens}{\otimes}

\newcommand{\be}{\begin{equation}}
\newcommand{\ee}{\end{equation}}
\newcommand{\bea}{\begin{eqnarray}}
\newcommand{\eea}{\end{eqnarray}}

\begin{document}

\title{On Unitary Evolution in Quantum Field Theory in Curved Spacetime}
\author{Daniele Colosi}\email{colosi@matmor.unam.mx}
\author{Robert Oeckl}\email{robert@matmor.unam.mx}
\affiliation{Instituto de Matemáticas, Universidad Nacional Autónoma de México,\\ Campus Morelia, C.P.~58190, Morelia, Michoacán, Mexico}

\begin{abstract}
We investigate the question of unitarity of evolution between hypersurfaces in quantum field theory in curved spacetime from the perspective of the general boundary formulation. Unitarity thus means unitarity of the quantum operator that maps the state space associated with one hypersurface to the state space associated with the other hypersurface. Working in Klein-Gordon theory, we find that such an evolution is generically unitary given a one-to-one correspondence between classical solutions in neighborhoods of the respective hypersurfaces. This covers the case of pairs of Cauchy hypersurfaces, but also certain cases where hypersurfaces are timelike. The tools we use are the Schrödinger representation and the Feynman path integral.

\end{abstract}

\maketitle

\section{Introduction}

Time evolution is implemented by unitary operators in quantum theory in order for a consistent probability interpretation to be possible, where probabilities are conserved in time. In non-relativistic quantum mechanics this is generally not an issue. Also in flat-spacetime quantum field theory, the unitarity of time evolution between equal-time hypersurfaces presents no problems. However, in quantum field theory it should also be possible to consider more general spacelike hypersurfaces and have unitary evolution between them. In curved spacetime such a more general setting becomes a necessity.

The first explicit description of evolution between general spacelike hypersurfaces in quantum field theory was given by Tomonaga \cite{Tom:relwave} and Schwinger \cite{Sch:qed1}. They proposed a functional differential equation, known as the Tomonaga-Schwinger equation that was to describe the infinitesimal evolution of wave functions between spacelike hypersurfaces. It has since been questioned not only whether such a heuristic equation can actually be made sense of, but also whether a unitary evolution between spacelike hypersurfaces can be implemented in principle \cite{ToVa:fevolqft}.

The purpose of the present article is to answer the latter question in the affirmative. What is more, we aim to show that unitary evolution actually occurs generically, at least in the free Klein-Gordon theory (possibly with source terms). Moreover, unitary ``evolution'' also occurs generically between certain classes of timelike hypersurfaces. In order to put this into perspective, it is necessary to clarify the conceptual framework on which we are building.

We shall adopt here the point of view furnished by the general boundary formulation \cite{Oe:boundary,Oe:GBQFT}, which has recently emerged as a viable description of quantum fields offering a new perspective on their dynamics \cite{Oe:timelike,Oe:KGtl,CoOe:spsmatrix,CoOe:smatrixgbf,CoOe:2deucl}. In particular, this means that a priori there is not any single Hilbert space of states for the quantum field theory. Instead, a Hilbert space of states is associated with each hypersurface in spacetime. In the present paper these will be leaves of a certain foliation. The states of these Hilbert spaces may be thought of as encoding information about physics near the associated hypersurface. Probability conservation and unitarity takes on a more general meaning than is usual \cite{Oe:GBQFT,Oe:probgbf}. Indeed, whenever the dynamics of the theory links states on two hypersurfaces in a one-to-one correspondence, the associated quantum operator should be unitary. In a quantization context this comes from a one-to-one correspondence between classical solutions near the hypersurfaces. Such a correspondence occurs of course in the standard case of spacelike Cauchy hypersurfaces. However, it may also occur for certain families of timelike hypersurfaces. Indeed, we show in the present article that given such a correspondence on the classical level (and certain additional assumptions, in particular about the metric) the corresponding quantum operator relating the two hypersurfaces is unitary. In such a setting a single Hilbert space can be recovered a posteriori, if desired, by identifying via these unitary operators all Hilbert spaces associated to hypersurfaces that are in correspondence with a given one.

The technical tools we shall use in this paper are quite simple and of limited sophistication. The point of view we are taking is naturally realized in the Schrödinger representation, where the state space associated with a hypersurface is simply the space of wave functions on field configurations on the hypersurface. Combining this with the Feynman path integral provides a convenient and sufficiently explicit description of quantum dynamics between hypersurfaces. In particular, this allows to avoid any infinitesimal treatment of hypersurface deformations as in the Tomonaga-Schwinger approach, which is potentially problematic.

We start in Section~\ref{sec:classical} with a treatment of the classical Klein-Gordon equation, finding the action in the relevant spacetime region in terms of boundary data. We proceed in Section~\ref{sec:quantum} to lay out the framework of the quantum theory, starting with the free theory and its unitarity in Section~\ref{sec:free}. Next follows a discussion of the compatibility of the vacuum wave function with the previous results in Section~\ref{sec:vacuum}. We then turn to a treatment of the free theory with a source term in Section~\ref{sec:source}, showing how the results generalize to this case. We proceed with a short discussion of the theory with perturbative interactions in Section~\ref{sec:perturbative}, without however trying to resolve the question of unitarity in this case. Finally in Section~\ref{sec:examples}, we illustrate our treatment with standard examples in Minkowski space. Throughout the paper, we assume for simplicity that certain fundamental operators commute, as happens in all examples that have been considered so far. The explicit form of some expressions in the more general case is contained in Appendix~\ref{sec:noncom}. In Appendix~\ref{sec:green} we show how some of the fundamental operators are expressed in terms of Green's functions.

\section{Classical theory}
\label{sec:classical}

Consider the free theory of a Klein-Gordon field $\phi$ with mass $m$ propagating on a four-dimensional Lorentzian spacetime with line element given by
\be
\xd s^2 = g_{\mu \nu} \xd x^{\mu} \xd x^{\nu}, \qquad \mu, \nu = 0,1,2,3.
\ee 
The action in a spacetime region $M$ is
\be
S_{M,0}(\phi) = \frac{1}{2} \int_M \xd^4 x \sqrt{|g|}\left( g^{\mu \nu} \partial_{\mu} \phi \, \partial_{\nu} \phi - m^2 \phi^2 \right),
\label{eq:action}
\ee
where the integration is extended over the region $M$ and we use the notation $\partial_{\mu}= \partial / \partial x^{\mu}$, and $g \equiv \det g_{\mu \nu}$. The equation of motion is obtained by varying the action with respect to the field and setting the variation equal to zero, yielding the Klein-Gordon equation
\be
\left(\frac{1}{\sqrt{|g|}} \partial_{\mu} \left(\sqrt{|g|} g^{\mu \nu} \partial_{\nu} \right) + m^2 \right) \phi(x)=0.
\label{eq:KG}
\ee
The action for a solution of the Klein-Gordon equation can be computed performing an integration by parts in (\ref{eq:action}) and using equation (\ref{eq:KG}), yielding
\be
S_M(\phi) = \frac{1}{2} \int_{\partial M} \xd^3 s \, \sqrt{|g^{(3)}|} \, n_{\mu}\, \phi \left(  g^{\mu \nu}  \partial_{\nu} \phi \right),
\label{eq:actreg}
\ee
where $s$ denotes some coordinates on the boundary $\partial M$ of $M$, $g^{(3)}$ is the determinant of the induced metric on the boundary and $n_{\mu}$ is the outward normal to $\partial M$.

From now on we suppose a smooth coordinate system $(t,\underline{x})$ which defines a foliation covering all or part of space-time. Thus, $\underline{x}\in\R^3$ are coordinates on the leaves while $t$ is a variable indexing the leaves. We suppose that $t\in I$, where $I$ might be a finite interval, a semi-infinite interval or all of $\R$. We identify $x^0=t$ and $x^i=\underline{x}^i$. We also require the metric to be block diagonal with respect to the foliation, i.e., $g^{0 i}=0=g^{i 0}$ for all $i\in\{1,2,3\}$. Note however, that in spite of the suggestive notation we do not necessarily require that $x^0$ be a temporal and $x^i$ be spatial coordinates.

We consider a spacetime region $M$ bounded by the disjoint union of two hypersurfaces that arise as leaves of the foliation, namely $\Sigma = \{(t,\underline{x}): t = \xi \}$ and $\hat{\Sigma} = \{(t,\underline{x}): t = \hat{\xi} \}$. We denote this spacetime region by $[\xi,\hat{\xi}]$. The action (\ref{eq:actreg}) evaluated on this region then takes the form,
\be
S_{[\xi,\hat{\xi}]}(\phi) = \frac{1}{2} \int \xd^3 \underline{x}\,
\left(\sqrt{|\hat{g}^{(3)}\hat{g}^{00}|} \phi(\hat{\xi},\underline{x})
 \left(\partial_{t} \phi \right)(\hat{\xi},\underline{x})
-\sqrt{|g^{(3)}g^{00}|} \phi(\xi,\underline{x})
 \left(\partial_{t} \phi \right)(\xi,\underline{x})\right),
\label{eq:actintreg}
\ee
where $g$ and $\hat{g}$ denote the metric restricted to the hypersurfaces $\Sigma$ and $\hat{\Sigma}$ respectively. We shall be interested in the value of $S_{[\xi,\hat{\xi}]}$ as a function of boundary field configurations
\be
\varphi(\underline{x})\defeq\phi(\xi,\underline{x}), \qquad \hat{\varphi}(\underline{x})\defeq\phi(\hat{\xi},\underline{x}) .
\ee
In order to obtain an explicit expression we will need to make further assumptions. In particular, we assume that solutions in $I\times\R^3$ give rise to a one-to-one correspondence between solutions in small neighborhoods of any two hypersurfaces with fixed $t\in I$. In the case where the coordinate system gives rise to a foliation by spacelike hypersurfaces this is the usual Cauchy property.
Fixing a particular way to specify ``initial'' data $\eta_a(\underline{x})$ and $\eta_b(\underline{x})$ we can write a solution of the Klein-Gordon as
\be
\phi(t,\underline{x}) = (X_a(t)\eta_a)(\underline{x}) + (X_b(t)\eta_b)(\underline{x}) ,
\label{eq:initsol}
\ee
where each $X_i(t)$ is a linear operator from the space of initial data $\eta_i$ to solutions evaluated on hypersurfaces at fixed values of $t$.
In the following we shall assume for simplicity that all these operators commute among each other and that certain operators satisfy a symmetry property. In all examples we have considered so far, this turns out to be the case. If the operators do not commute or the symmetry property is not satisfied, the computations become somewhat more involved, see Appendix~\ref{sec:noncom}.
Inverting the matrix operator equation
\be
 \begin{pmatrix}\varphi \\ \hat{\varphi}\end{pmatrix}
 =\begin{pmatrix}X_a(\xi) & X_b(\xi)\\ X_a(\hat{\xi}) & X_b(\hat{\xi})\end{pmatrix}
 \begin{pmatrix}\eta_a \\ \eta_b\end{pmatrix}
\ee
we obtain,\footnote{The boundary value problem we are considering here does of course in general not have a unique solution. Indeed, it would usually have infinitely many solutions. Thus, there are arbitrary choices involved in the inversion process we are describing. However, any such choice yields the same result in the moment we evaluate the action, i.e., once we arrive at (\ref{eq:actev}). We therefore allow ourselves to gloss over this detail as we did on previous occasions \cite{Oe:KGtl,CoOe:smatrixgbf} when we showed that standard results are correctly reproduced.}
\be
 \begin{pmatrix}\eta_a \\ \eta_b\end{pmatrix}=\frac{1}{\Delta(\xi,\hat{\xi})}
  \begin{pmatrix}X_b(\hat{\xi}) & -X_b(\xi)\\ -X_a(\hat{\xi}) & X_a(\xi)\end{pmatrix}
\begin{pmatrix}\varphi \\ \hat{\varphi}\end{pmatrix},
\ee
where $\Delta(\xi,\hat{\xi})\defeq X_a(\xi) X_b(\hat{\xi}) - X_a(\hat{\xi}) X_b(\xi)$. Reinserting this into (\ref{eq:initsol}) yields the solution in $I\times\R^3$ as a function of the boundary data,
\be
 \phi(t,\underline{x}) 
= \left(\frac{\Delta(t, \hat{\xi})}{\Delta(\xi, \hat{\xi})} \varphi\right)(\underline{x})
 +\left(\frac{\Delta(\xi, t)}{\Delta(\xi, \hat{\xi})} \hat{\varphi}\right)(\underline{x}) .
\label{eq:solbdyval}
\ee
This allows to evaluate (\ref{eq:actintreg}), resulting in,
\be
S_{[\xi,\hat{\xi}]}(\phi)
=  \frac{1}{2} \int \xd^3 \underline{x} \, 
\begin{pmatrix}\varphi & \hat{\varphi} \end{pmatrix} W_{[\xi,\hat{\xi}]}  
\begin{pmatrix} \varphi \\ \hat{\varphi} \end{pmatrix},
\label{eq:actev}
\ee
where the $W_{[\xi,\hat{\xi}]}$ is a 2x2 matrix with elements $W_{[\xi,\hat{\xi}]}^{(i,j)}, (i,j=1,2),$ given by
\begin{align}
W_{[\xi,\hat{\xi}]}^{(1,1)} =& - \sqrt{|g^{(3)}g^{0 0}|} \,
  \frac{\Delta_1(\xi, \hat{\xi})}{\Delta(\xi, \hat{\xi})},
& W_{[\xi,\hat{\xi}]}^{(1,2)} =& - \sqrt{|g^{(3)}g^{0 0}|} \,
  \frac{\Delta_2(\xi, \xi)}{\Delta(\xi, \hat{\xi})}, \nonumber \\
W_{[\xi,\hat{\xi}]}^{(2,1)} =& \sqrt{|\hat{g}^{(3)}\hat{g}^{0 0}|} \,
  \frac{\Delta_1(\hat{\xi}, \hat{\xi})}{\Delta(\xi, \hat{\xi})},
& W_{[\xi,\hat{\xi}]}^{(2,2)} =& \sqrt{|\hat{g}^{(3)}\hat{g}^{0 0}|}\,
  \frac{\Delta_2(\xi, \hat{\xi})}{\Delta(\xi, \hat{\xi})},
\label{eq:W2}
\end{align}
where
\be
 \Delta_1(\xi,\hat{\xi})\defeq\partial_t \Delta(t,\hat{\xi})\big|_{t =\xi}\qquad
 \Delta_2(\xi,\hat{\xi})\defeq\partial_t \Delta(\xi,t)\big|_{t =\hat{\xi}} .
\ee

The space of smooth solutions of (\ref{eq:KG}) on $\R^3\times I$ is equipped with the following symplectic form
\be
\Omega(\phi_1, \phi_2) = \frac{1}{2}\int_{\Sigma} \xd^3 \underline{x}  \, \sqrt{|g^{(3)} g^{0 0}|} \left( \phi_1 \, \partial_t \phi_2 - \phi_2 \, \partial_t \phi_1 \right),
\label{eq:sympl}
\ee
which is independent of the choice of leaf $\Sigma$ of the foliation. (See \cite{Woo:geomquant}. In case of the leaves being spacelike, this is just the standard symplectic form.) This implies that the operator
\be
{\mathcal{W}} := 
\sqrt{|g^{(3)} g^{0 0}|}\, \Delta_2(t,t)=-\sqrt{|g^{(3)} g^{0 0}|}\, \Delta_1(t,t)
\ee
is independent of $t$. Note that this immediately implies $W_{[\xi,\hat{\xi}]}^{(1,2)}=W_{[\xi,\hat{\xi}]}^{(2,1)}$.

\section{Quantum theory and unitarity}
\label{sec:quantum}

The passage to the quantum theory is implemented by the Feynman path integral prescription. Moreover, the quantum dynamics of the field is described in the Schrödinger representation, where the quantum states are wave functionals on the space of field configurations.
Thus, with a given spacetime hypersurface $\Sigma$ we associate the space of state $\cH_\Sigma$ of wave functions of field configurations on $\Sigma$. This state space carries the following inner product,
\be
\langle \psi_{\Sigma}, \psi_{\Sigma}'\rangle \defeq \int \xD \varphi \, \overline{\psi_{\Sigma}(\varphi)} \, \psi_{\Sigma}'(\varphi),
\label{eq:inner-prod}
\ee
where the integral is over all field configurations $\varphi$ on the $\Sigma$.
Amplitudes $\rho_{M}:\cH_{\partial M}\to\C$ are associated to spacetime regions $M$. State spaces and amplitudes satisfy a number of consistency conditions, see \cite{Oe:GBQFT} or \cite{Oe:KGtl}.

Consider now as above regions $[\xi,\hat{\xi}]$ that are bounded by hypersurfaces $\Sigma$ at $\xi$ and $\hat{\Sigma}$ at $\hat{\xi}$. In this case, the field propagator associated with the spacetime region is formally defined as
\be
Z_{[\xi,\hat{\xi}]}(\varphi, \hat{\varphi}) = \int_{\phi|_{\Sigma}=\varphi, \, \phi|_{\hat{\Sigma}}=\hat{\varphi} } \xD\phi\, e^{\im S_{[\xi,\hat{\xi}]}(\phi)},
\label{eq:proppint}
\ee
where $S_{[\xi,\hat{\xi}]}(\phi)$ is the action of the field in the region $[\xi,\hat{\xi}]$ and the integration is extended over all field configurations $\phi$ that reduce to the boundary configurations $\varphi$ and $\hat{\varphi}$ on the boundary hypersurfaces $\Sigma$ and $\hat{\Sigma}$ respectively. All the information on the dynamical evolution of the field between boundary configurations $\varphi$ and $\hat{\varphi}$ is encoded in the propagator (\ref{eq:proppint}).
The amplitude associated with the region $[\xi,\hat{\xi}]$ and a state $\psi_\xi\tens\overline{\psi_{\hat{\xi}}}\in\cH_\xi\tens\cH_{\hat{\xi}}^*=\cH_{\partial [\xi,\hat{\xi}]}$
is then
\be
\rho_M(\psi_{\xi} \otimes \overline{\psi_{\hat{\xi}}})
 = \int \xd \varphi \, \xd \hat{\varphi} \, \psi_{\xi}(\varphi) \, \overline{\psi_{\hat{\xi}} (\hat{\varphi})}\, Z_{[\xi,\hat{\xi}]}(\varphi, \hat{\varphi}).
\ee
The main consistency condition for amplitudes is the composition property. In the present context this property means that the composition of two amplitudes for evolving in the $t$ variable first from $\xi_1$ to $\xi_2$ and then from $\xi_2$ to $\xi_3$ equals the amplitude for the direct evolution from $\xi_1$ to $\xi_3$. This translates into the following identity for propagators:
\be
 Z_{[\xi_1,\xi_3]}(\varphi_1,\varphi_3)=
 \int \xD\varphi_2\, Z_{[\xi_1,\xi_2]}(\varphi_1,\varphi_2)
  Z_{[\xi_2,\xi_3]}(\varphi_2,\varphi_3)  .
\label{eq:composition}
\ee

We recall from \cite{Oe:GBQFT} or \cite{Oe:KGtl} that unitarity of the quantum evolution between hypersurfaces $\Sigma$ (at $t=\xi$) and $\hat{\Sigma}$ (at $t=\hat{\xi}$) is equivalent to the equation
\be
\int \xd \hat{\varphi} \, \overline{Z_{[\xi,\hat{\xi}]}(\varphi,\hat{\varphi})} \,
 Z_{[\xi,\hat{\xi}]}(\varphi',\hat{\varphi}) = \delta(\varphi - \varphi'),
\label{eq:unitarity}
\ee
where $\varphi,\varphi'$ are field configurations on the hypersurface $\Sigma$ and $\hat{\varphi}$ are field configurations on the hypersurface $\hat{\Sigma}$ that are integrated over. The delta function is meant with respect to the integral over field configurations on $\Sigma$. As is easy to see, this equation precisely guarantees that the inner product (\ref{eq:inner-prod}) of two states at $t=\xi$ remains the same when each states is evolved via the propagator $Z_{[\xi,\hat{\xi}]}$ from $t=\xi$ to $t=\hat{\xi}$.

\subsection{Free Theory}
\label{sec:free}

The first case we are interested in is the free theory with the action given by (\ref{eq:action}). We shall use a subscript $0$ in the relevant quantities to indicate that we are referring to this case. The propagator (\ref{eq:proppint}) can be evaluated by shifting the integration variable by a classical solution, $\phi_\text{cl}$, matching the boundary configurations in $\partial [\xi,\hat{\xi}] = \Sigma \cup \hat{\Sigma}$, i.e. $\phi_\text{cl}|_{\Sigma} = \varphi$ and $\phi_\text{cl}|_{\hat{\Sigma}} = \hat{\varphi}$. Explicitly,
\be
Z_{[\xi,\hat{\xi}],0}(\varphi, \hat{\varphi}) = \int_{\phi|_{\Sigma}=\varphi, \, \phi|_{\hat{\Sigma}}=\hat{\varphi} } \xD\phi\, e^{\im S_{[\xi,\hat{\xi}],0}(\phi)}
=  \int_{\phi|_{\partial [\xi,\hat{\xi}]}=0} \xD\phi\,
 e^{\im  S_{[\xi,\hat{\xi}],0}(\phi_{cl}+\phi)}
 = N_{[\xi,\hat{\xi}],0} \,  e^{\im  S_{[\xi,\hat{\xi}],0}(\phi_\text{cl})} ,
\label{eq:prop0}
\ee
where the normalization factor is formally given by
\be
N_{[\xi,\hat{\xi}],0}=\int_{\phi|_{\partial [\xi,\hat{\xi}]}=0}
 \xD\phi\, e^{\im S_{[\xi,\hat{\xi}],0}(\phi)}.
\label{eq:n00}
\ee
Explicit calculation shows that the composition property (\ref{eq:composition}) of the propagators (\ref{eq:prop0}) is satisfied if the following identity for the normalization factors (\ref{eq:n00}) holds,
\be
 N_{[\xi_1,\xi_3],0} = N_{[\xi_1,\xi_2],0} N_{[\xi_2,\xi_3],0}
 \int\xD\varphi_2\, \exp\left(-\frac{\im}{2}
 \int\xd^3 x\,  \varphi_2 \frac{W^{(1,2)}_{[\xi_1,\xi_2]}
 W^{(1,2)}_{[\xi_2,\xi_3]}}{W^{(1,2)}_{[\xi_1,\xi_3]}} \varphi_2\right) .
\ee
Using the identity
\be
 \int\xD\varphi\,\exp\left(-\frac{1}{2}\int \xd^3 x\,\varphi A \varphi\right)
 =\det\left(\frac{A}{2\pi}\right)^{-\frac{1}{2}},
\ee
we obtain
\be
 N_{[\xi_1,\xi_3],0} N_{[\xi_1,\xi_2],0}^{-1} N_{[\xi_2,\xi_3],0}^{-1}
 =\det\left(\left(\frac{\im W^{(1,2)}_{[\xi_1,\xi_3]}}{2\pi}\right)^{-1}
 \frac{\im W^{(1,2)}_{[\xi_1,\xi_2]}}{2\pi}\frac{\im W^{(1,2)}_{[\xi_2,\xi_3]}}{2\pi}\right)^{-\frac{1}{2}} .
\ee
This suggests the following solution for the normalization factor, which we shall henceforth adopt,
\be
 N_{[\xi,\hat{\xi}],0}=\det\left(\frac{\im W^{(1,2)}_{[\xi,\hat{\xi}]}}{2\pi}\right)^{\frac{1}{2}} .
\label{eq:normf}
\ee

We turn to the unitarity condition (\ref{eq:unitarity}). Since the action is real we have,
\[
 \overline{Z_{[\xi,\hat{\xi}],0}(\varphi,\hat{\varphi})}
=\overline{N_{[\xi,\hat{\xi}],0}} e^{-\im S_{[\xi,\hat{\xi}],0}(\phi_\text{cl})} .
\]
Thus,
\begin{align}
\int \xd \hat{\varphi} \, \overline{Z_{[\xi,\hat{\xi}],0}(\varphi,\hat{\varphi})} \,
 Z_{[\xi,\hat{\xi}],0}(\varphi',\hat{\varphi})
 &= \left|N_{[\xi,\hat{\xi}],0}\right|^2 \int \xD \hat{\varphi} \,
 \exp \left(\frac{\im}{2} \int \xd^3 x \left[ \hat{\varphi} \, 2 W_{[\xi,\hat{\xi}]}^{(1,2)} \left( \varphi' - \varphi \right)  - \varphi W_{[\xi,\hat{\xi}]}^{(1,1)} \varphi + \varphi' W_{[\xi,\hat{\xi}]}^{(1,1)} \varphi' \right] \right) \nonumber\\
&= \left|N_{[\xi,\hat{\xi}],0}\right|^2 \exp \left(\frac{\im}{2} \int \xd^3 x \left[ \varphi' W_{[\xi,\hat{\xi}]}^{(1,1)} \varphi'- \varphi W_{[\xi,\hat{\xi}]}^{(1,1)} \varphi \right] \right)
\delta\left(W_{[\xi,\hat{\xi}]}^{(1,2)} (\varphi'-\varphi) \right) \nonumber\\
&= \left|N_{[\xi,\hat{\xi}],0}\right|^2 \det \left( \frac{|W_{[\xi,\hat{\xi}]}^{(1,2)}|}{2 \pi} \right)^{-1} \delta(\varphi'-\varphi) \nonumber\\
&= \delta(\varphi'-\varphi),
\end{align}
where we used the normalization factor $N_{[\xi,\hat{\xi}],0}$ as found in (\ref{eq:normf}).

\subsection{Compatibility of the vacuum}
\label{sec:vacuum}

In the following we are going to verify that a notion of vacuum, if it exists, is compatible with the structures we have found for the free theory, including the unitarity condition. According to the axioms of the general boundary formulation \cite{Oe:GBQFT}, a vacuum consists of a particular state associated with each leaf of the foliation. These vacuum states have to be related by evolution between leaves. In the free theory, this amounts to
\be
\psi_{\hat{\xi},0}(\hat{\varphi}) = \int \xD \varphi \, \psi_{\xi,0} (\varphi) \, Z_{[\xi, \hat{\xi}],0}(\varphi, \hat{\varphi}). 
\ee
Making a Gaussian ansatz for the form of the vacuum state,
\be
\psi_{\xi,0}(\varphi) = C_{\xi} \exp \left(- \frac{1}{2} \int \xd ^3 \underline{x} \, \varphi(\underline{x}) (A_{\xi} \varphi)(\underline{x}) \right),
\ee
where $A_\xi$ is some operator and
$C_{\xi}$ is a normalization factor ensuring that the vacuum state is normalized, via
\be
|C_{\xi}|^2 = \det \left[ \frac{A_{\xi} + \overline{A_{\xi}}}{2 \pi}\right]^\frac{1}{2}.
\label{eq:vacnorm}
\ee
Evolving the vacuum state from $t=\xi$ to $t=\hat{\xi}$ with the field propagator yields the consistency condition,
\be
C_{\hat{\xi}}= N_{[\xi, \hat{\xi}],0} \, C_{\xi} \, \det \left[ \frac{A_{\xi} - \im W_{[\xi, \hat{\xi}]}^{(1,1)}}{2 \pi} \right]^{-1/2}.
\label{eq:N_M}
\ee
As we shall see, this equation can indeed be satisfied and allows us to fix relative phases for the normalization factors $C_t$. The general form of the operator $A_t$ has been derived in \cite{Col:vacuum},
\be
 A_t=-\im \sqrt{|g^{(3)}g^{0 0}|}\,\frac{\partial_t (c_a X_a(t)+c_b X_b(t))}{c_a X_a(t)+c_b X_b(t)} ,
\ee
where $c_a$ and $c_b$ are complex numbers, characterizing the vacuum wave function. In order for the vacuum state to be normalizable, these complex numbers must satisfy the condition $\overline{c_a} c_b -\overline{c_b} c_a\neq 0$. One easily checks that
\be
 C_t=\det\left[\frac{\sqrt{|g^{(3)}g^{0 0}|}\, \im(\overline{c_a} c_b -\overline{c_b} c_a)\Delta_1(t,t)}{2\pi (c_a X_a(t)+c_b X_b(t))^2}\right]^{\frac{1}{4}}
\ee
provides a solution to both (\ref{eq:vacnorm}) and (\ref{eq:N_M}), where equation (\ref{eq:normf}) has been used.

\subsection{Interaction with a source field}
\label{sec:source}

We consider in this section the interaction with a real source field $\mu$. The action takes the form
\be
S_{M,\mu}(\phi) = S_{M,0}(\phi) + \int_M \xd ^4 x \, \sqrt{|g|} \, \mu(x) \phi(x) .
\ee
As before we are interested in the field propagator in the region $[\xi,\hat{\xi}]$ where we can write the action in terms of boundary field configurations as
\be
S_{[\xi,\hat{\xi}],\mu}(\phi) = S_{[\xi,\hat{\xi}],0}(\phi) + \int \xd ^3 \underline{x}  \left( \mu_{\xi}( \underline{x}) \varphi( \underline{x}) + \mu_{\hat{\xi}}( \underline{x}) \hat{\varphi}( \underline{x}) \right),
\ee
with
\begin{align}
\mu_{\xi}(\underline{x}) &\defeq \int_\xi^{\hat{\xi}} \xd t \, \sqrt{|g|} \, \frac{\Delta(t, \hat{\xi})}{\Delta(\xi, \hat{\xi})} \, \mu(t, \underline{x}), \\
\mu_{\hat{\xi}}(\underline{x}) &\defeq \int_\xi^{\hat{\xi}} \xd t \, \sqrt{|g|} \, \frac{\Delta(\xi, t)}{\Delta(\xi, \hat{\xi})} \, \mu(t, \underline{x}).
\end{align}
We can then write the corresponding propagator as
\be
Z_{[\xi,\hat{\xi}],\mu}(\varphi, \hat{\varphi}) = \frac{N_{[\xi,\hat{\xi}],\mu}}{N_{[\xi,\hat{\xi}],0}} \, Z_{[\xi,\hat{\xi}],0}(\varphi, \hat{\varphi}) \, 
\exp \left( \im \int \xd ^3 \underline{x} \left( \mu_{\xi}( \underline{x}) \varphi( \underline{x}) + \mu_{\hat{\xi}}( \underline{x}) \hat{\varphi}( \underline{x}) \right) \right).
\label{eq:propsrc}
\ee
The normalization factor can be expressed as
\be
\frac{N_{[\xi,\hat{\xi}],\mu}}{N_{[\xi,\hat{\xi}],0}} = \exp \left( \frac{\im}{2} \int_{[\xi,\hat{\xi}]} \xd^4 x \, \sqrt{|g|} \, \mu(x) \, \alpha(x) \right),
\label{eq:normsrc}
\ee
where $\alpha$ is a solution of the inhomogeneous Klein-Gordon equation
\be
(\Box + m^2) \alpha =\mu,
\label{eq:KGin}
\ee
with boundary conditions,
\be
\alpha |_{t=\xi} = \alpha |_{t=\hat{\xi}}=0.
\ee
$\alpha$ can be written as
\be
\alpha(t, \underline{x}) = -\int_{\xi}^{\hat{\xi}} \xd t' \sqrt{|g(t',\underline{x})|} \left(\theta(t'-t) \frac{\Delta(\xi,t)\Delta(t',\hat{\xi})}{\mathcal{W}\Delta(\xi,\hat{\xi})} + \theta(t-t') \frac{\Delta(\xi,t')\Delta(t,\hat{\xi})}{\mathcal{W}\Delta(\xi,\hat{\xi})} \right) \mu(t', \underline{x}) .
\label{eq:alpha}
\ee
Given these ingredients one may explicitly verify that (\ref{eq:propsrc}) satisfies the composition property.

Checking the unitarity condition (\ref{eq:unitarity}) yields,
\begin{align}
\int \xd \hat{\varphi} \, \overline{Z_{[\xi, \hat{\xi}], \mu} (\varphi, \hat{\varphi})} \, Z_{[\xi, \hat{\xi}], \mu} (\varphi', \hat{\varphi}) 
&= \left| \frac{N_{[\xi,\hat{\xi}],\mu}}{N_{[\xi,\hat{\xi}],0}} \right|^2
\int \xd \hat{\varphi} \, \overline{Z_{[\xi, \hat{\xi}], 0} (\varphi, \hat{\varphi})} \, Z_{[\xi, \hat{\xi}], 0} (\varphi', \hat{\varphi}) 
\exp \left( \im \int \xd ^3 \underline{x} \, \mu_{\xi}( \underline{x}) \left( \varphi'( \underline{x}) - \varphi( \underline{x}) \right)  \right)
, \nonumber\\
&= \left| \frac{N_{[\xi,\hat{\xi}],\mu}}{N_{[\xi,\hat{\xi}],0}} \right|^2 \,\exp \left( \im \int \xd ^3 \underline{x} \, \mu_{\xi}( \underline{x}) \left( \varphi'( \underline{x}) - \varphi( \underline{x}) \right)  \right)\,
 \delta(\varphi - \varphi'), \\
&= \delta(\varphi - \varphi') .
\end{align}
In the last step we have used that the quotient (\ref{eq:normsrc}) has modulus one since the integrand on the right hand side of (\ref{eq:normsrc}) is real. This confirms unitarity in the presence of a source field.

\subsection{Perturbative interaction}
\label{sec:perturbative}

Finally, we consider the perturbatively interacting theory described by the action
\be
 S_{M,V}(\phi)=S_{M,0}(\phi)+\int_M \xd^4 x\sqrt{|g(x)|}\, V(x,\phi(x)),
\ee
where $V$ is an arbitrary potential. We may write this using functional derivatives as,
\be
\exp \left( \im S_{M,V}(\phi)\right) = \exp \left(\im \int_M \xd^4 x\sqrt{|g(x)|} \, V\left(x, -\im \frac{\delta}{\delta \mu(x)} \right) \right) \exp \left( \im S_{M,\mu}(\phi)\right) \bigg|_{\mu=0}.
\ee
The propagator for the spacetime region $[\xi,\hat{\xi}]$ can then be written as
\begin{align}
Z_{[\xi,\hat{\xi}],V}(\varphi, \hat{\varphi})
=& \exp \left(\im \int_{[\xi,\hat{\xi}]} \xd^4 x \, V\left(x, -\im \frac{\delta}{\delta \mu(x)} \right) \right) Z_{[\xi,\hat{\xi}],\mu}(\varphi, \hat{\varphi}) \bigg|_{\mu=0}, \nonumber\\
=& Z_{[\xi,\hat{\xi}],0}(\varphi, \hat{\varphi})
\exp \left(\im \int_{[\xi,\hat{\xi}]} \xd^4 x \, V\left(x, -\im \frac{\delta}{\delta \mu(x)} \right) \right)
\frac{N_{[\xi,\hat{\xi}],\mu}}{N_{[\xi,\hat{\xi}],0}} \, 
\exp \left( \im \int_{[\xi,\hat{\xi}]} \xd^4 x \sqrt{|g(x)|} \,  \mu(x) \, \phi(x) \right) \bigg|_{\mu=0}, \nonumber\\
=& Z_{[\xi,\hat{\xi}],0}(\varphi, \hat{\varphi})
\exp \left(\im \int_{[\xi,\hat{\xi}]} \xd^4 x \, V\left(x, -\im \frac{\delta}{\delta \mu(x)} \right) \right)
\exp \left( \frac{\im}{2} \int_{[\xi,\hat{\xi}]} \xd^4 x  \sqrt{|g(x) |} \, \mu(x) \, \alpha_\mu(x) \right) 
\nonumber\\
& \times
\exp \left( \im \int_{[\xi,\hat{\xi}]} \xd^4 x \sqrt{|g(x)|} \,  \mu(x) \, \phi(x) \right) \bigg|_{\mu=0},
\label{eq:propgenint}
\end{align}
where in the last step the quotient of normalization factors (\ref{eq:normsrc}) has been substituted, and the function $\alpha_\mu$ is given by (\ref{eq:alpha}), where the subscript is meant to emphasize that $\alpha_\mu$ depends on $\mu$.
The use of this functional derivative technique to express the field propagator in the presence of a perturbative interaction turns out to be inadequate to study the unitarity with the method proposed in this work, namely via the composition of the propagators (\ref{eq:unitarity}). An alternative way to analyze the evolution for a general interacting theory would be to study the properties of the S-matrix. We shall elaborate on this elsewhere.

\section{Examples}
\label{sec:examples}

Although the present article is written very much in order to provide a context for unitary evolution in curved spacetime, we limit ourselves in this example section to flat Minkowski spacetime. However, we emphasize instead the fact that the ``evolution'' does not have to be temporal. Thus, after considering the standard setting of evolution between equal-time hypersurfaces we switch to a unitary ``evolution'' in a radial direction as considered in \cite{Oe:KGtl,CoOe:spsmatrix,CoOe:smatrixgbf}. We do not explicitly repeat the calculations involved in showing the composition property, unitarity etc., but merely indicate the involved operators and normalization factors.

\subsection{Time-interval region in Minkowski space}

The foliation of Minkowski spacetime is the standard one here, indexed by a global time coordinate $t\in\R$, while the leaves of the foliation have coordinates $\underline{x}\in\R^3$. The initial data hypersurface is located at $t=0$. We exemplify two different choices for initial data. If we choose $(\eta_a,\eta_b)$ to be the positive and negative energy components of the field at time $t=0$ we have
\begin{equation}
 X_a(t)=e^{-\im \omega t},\quad X_b(t)=e^{\im \omega t}\quad\text{with}\quad
 \omega \defeq \sqrt{- \Delta_{\underline{x}} + m^2},
\end{equation}
$\Delta_{\underline{x}}$ being the Laplacian in the coordinates ${\underline{x}}$.
If we choose $(\eta_a,\eta_b)$ instead to represent the initial value and its temporal derivative at time $t=0$ we have instead,
\begin{equation}
 X_a(t)=\cos(\im \omega t),\quad X_b(t)=\frac{\sin(\im \omega t)}{\im\omega}.
\end{equation}
For a time interval $[t_1,t_2]$ the matrix $W_{[t_1,t_2]}$ of (\ref{eq:W2}) results to be
\begin{equation}
W_{[t_1,t_2]} = \frac{\omega}{ \sin \omega (t_2-t_1)}
\begin{pmatrix}  \cos \omega (t_2-t_1) & -1 \\ 
-1 & \cos \omega (t_2-t_1) \end{pmatrix},
\end{equation}
while the normalization factor $N_{[t_1,t_2]}$ of (\ref{eq:normf}) turns out to be
\be
N_{[t_1,t_2]} = \det \left( \frac{\im \omega}{2 \pi \, \sin \omega (t_2-t_1)} \right)^{\frac{1}{2}} .
\label{eq:normf1}
\ee

\subsection{Hypercylinder in Minkowski space}

We consider now a different foliation of Minkowski space, see \cite{Oe:KGtl,CoOe:smatrixgbf} for details. We use coordinates $(t,r,\Omega)$ where $t$ is the usual time coordinate, while $r$ and $\Omega$ are spherical coordinates in space. ($\Omega$ is a collective coordinate for the angles on the $2$-sphere.) The leaves of the foliation are now the timelike hypersurfaces of constant $r$, where $r\in(0,\infty)$. (We are missing the time axis in this foliation.) Note that the Minkowski metric diagonalizes in the required way.
In the present context it makes sense to divide the classical solutions into those that converge when $r\to 0$ and those that diverge when $r\to 0$. Using $\eta_a$ to encode initial data of solutions that converge and $\eta_b$ for solutions that diverge, we can define
\begin{equation}
 X_a(r)\, e^{\im E t} Y_{l}^m(\Omega)\defeq a_l(E,r) e^{\im E t} Y_{l}^m(\Omega),\quad
 X_b(r)\, e^{\im E t} Y_{l}^m(\Omega)\defeq b_l(E,r) e^{\im E t} Y_{l}^m(\Omega),
\end{equation}
where $Y_l^m$ are the usual spherical harmonics, $E\in\R$ and $a_l$ and $b_l$ are defined as
\bea
a_l(E, r) =\left\{ \begin{matrix} j_l(r \sqrt{E^2 - m^2}), & \mbox{if} \ \ E^2>m^2, \\ i^+_l(r \sqrt{m^2 -E^2}), & \mbox{if} \ \  E^2<m^2, \end{matrix} \right.
\ \ \mbox{and} \ \
b_l(E, r) =\left\{ \begin{matrix} n_l(r \sqrt{E^2 - m^2}), & \mbox{if} \ \ E^2>m^2, \\ i^-_l(r \sqrt{m^2 -E^2}), & \mbox{if} \ \  E^2<m^2, \end{matrix} \right.
\eea
where $j_l, n_l, i^+_l$ and $i^-_l$ are the spherical Bessel functions of the first and second kind and the modified spherical Bessel functions of the first and second kind respectively.
Given $0<R<\hat{R}$, the matrix $W_{[R,\hat{R}]}$ of (\ref{eq:W2}) is the operator valued 2x2 matrix
\be
W_{[R,\hat{R}]} = \frac{1}{ \delta_l(E,R,\hat{R})}
\begin{pmatrix}  - R^2 \sigma_l(E,\hat{R},R) & 1/p \\ 
1/p & -\hat{R}^2 \sigma_l(E,R,\hat{R}) \end{pmatrix},
\ee
where 
\be
p := \left\{ \begin{matrix}  \sqrt{E^2 - m^2}, & \mbox{if} \ \ E^2>m^2, \\ \im \sqrt{m^2 -E^2}, & \mbox{if} \ \  E^2<m^2, \end{matrix} \right.
\ee
and the functions $\delta_l$ and $\sigma_l$ are to be understood as operators defined as
\be
\delta_l(E,R,\hat{R})= a_l(E,R) b_l(E, \hat{R}) - b_l(E,R) a_l(E, \hat{R}), \qquad
\sigma_l(E,R,\hat{R})= a_l(E,R) b_l'(E, \hat{R}) - b_l(E,R) a_l'(E, \hat{R}),
\ee
where the prime indicates derivative with respect to the second argument. The normalization factor $N_{[R,\hat{R}]}$ of (\ref{eq:normf}) is
\be
N_{[R,\hat{R}]} = \det \left( \frac{\im }{2 \pi \, p \delta_l(E,R,\hat{R})} \right)^{\frac{1}{2}}.
\label{eq:normfR}
\ee

\section{Conclusions}

In the present paper we have investigated the quantum evolution of Schrödinger wave functions of Klein-Gordon theory along special foliations of (a part of) curved spacetime. We have worked out the field propagators between leaves of the foliation and found that they yield unitary ``evolution'' operators between the corresponding spaces of wave functions. The key required properties of the foliation were two: (a) There has to be a one-to-one correspondence between classical solutions in a neighborhood of any pair of leaves. (b) The spacetime metric has to take a block diagonal form with respect to the foliation. These conditions can be naturally realized for Cauchy hypersurfaces in globally hyperbolic manifolds. However, as we have shown, there are also interesting examples involving unitary ``evolution'' between timelike hypersurfaces. As already hinted at in the introduction, this implements probability conservation ``in space'' rather than in time, see \cite{Oe:GBQFT}.

Although we have used foliations for convenience, it is clear from our construction that the amplitude (and thus the corresponding ``evolution'' operator) for any pair of hypersurfaces is manifestly independent of the rest of the foliation. The construction of the field propagator that determines the amplitude did not involve information about the intermediate foliation. Indeed, no such intermediate foliation needs to exist. (However, condition (a) above needs to be satisfied for the pair of hypersurfaces and condition (b) needs to be satisfied in neighborhoods of the hypersurfaces.)

It is now time to clarify differences to other approaches at unitary evolution in curved spacetime that have run into difficulties. As already mentioned in the introduction the key difference to the original approach of Tomonaga and Schwinger is more of a technical than of a conceptual nature. The problems inherent in describing hypersurface deformations in terms of infinitesimal generators and the ambiguities associated to ``integrating'' these generators are simply avoided in a Feynman path integral approach. Instead of seeking a generalization of the Schrödinger equation we directly look for a generalization of transition amplitudes.

Another approach that has been followed in the literature is to try to generalize the fact that the Poincaré group acts on the state space of quantum field theory in Minkowski space. Of course, a generic Lorentzian manifold will not have any isometries. However, supposing there is a Killing vector field leaving the Lagrangian invariant, this will induce an action on the phase space of the classical field theory by symplectic transformations. One may then hope to implement such an action on a Hilbert space for the quantum field theory and interpret it as the generator of evolutions along the vector field \cite{Hel:stressenergy,Hel:hamscal}. Indeed, one may even try actions coming from suitable more general (non-isometric) spacetime transformations \cite{ToVa:fevolqft}. In either case, serious difficulties or even no-go theorems have resulted. Let us emphasize that in our approach no such action is obtained. Indeed, the only natural way to relate the Hilbert spaces associated with different hypersurfaces in our approach is precisely via the unitary operators representing evolution between them. Absent any other way to relate the Hilbert spaces, no non-trivial action can be constructed.

A deficiency of our paper may be seen in the fact that the concrete examples we have presented concern only Minkowski spacetime, rather than the curved spacetimes our treatment is mainly aimed at. However, some curved spacetime examples have already been partially worked out in the present setting \cite{Col:vacuum,Col:desitterletter} and others are to follow.

\appendix

\section{Non-commuting operators}
\label{sec:noncom}

In the main body of the article, we have assumed that all operators of the form $X_a(t), X_b(t')$ commute and that the operators $W_{[\xi,\hat{\xi}]}^{(i,j)}$ satisfy a symmetry condition of the form
\be
 \int \xd^3 \underline{x}\, \varphi(\underline{x}) (W_{[\xi,\hat{\xi}]}^{(i,j)} \varphi')(\underline{x})
 =  \int \xd^3 \underline{x}\, \varphi'(\underline{x}) (W_{[\xi,\hat{\xi}]}^{(i,j)} \varphi)(\underline{x}),
\ee
where $\varphi,\varphi'$ are field configuration data. If this is not the case, the computations become somewhat more involved. In the following we will present merely the resulting elements of the matrix $W_{[\xi,\hat{\xi}]}$ that reduce in the commutative case to the expressions given in (\ref{eq:W2}). The general result can be written as follows:
\begin{align}
W_{[\xi,\hat{\xi}]}^{(1,1)} & = -\sqrt{|g^{(3)} g^{0 0}|}
\left(X_a' X_a^{-1} D^{-1} \hat{X}_b X_b^{-1}
  -X_b' X_b^{-1}D^{-1} \hat{X}_a X_a^{-1}\right),\\
W_{[\xi,\hat{\xi}]}^{(1,2)} & =-\sqrt{|g^{(3)} g^{0 0}|}\left( X_b' X_b^{-1}D^{-1}
 - X_a' X_a^{-1}D^{-1}\right),\\
W_{[\xi,\hat{\xi}]}^{(2,1)} & =\sqrt{|\hat{g}^{(3)} \hat{g}^{0 0}|}
  \left( \hat{X}_a' X_a^{-1} D^{-1} \hat{X}_b X_b^{-1}
  -\hat{X}_b' X_b^{-1} D^{-1} \hat{X}_a X_a^{-1}\right),\\
W_{[\xi,\hat{\xi}]}^{(2,2)} & =\sqrt{|\hat{g}^{(3)} \hat{g}^{0 0}|}\left( \hat{X}_b' X_b^{-1} D^{-1}
  - \hat{X}_a' X_a^{-1}D^{-1}\right).
\end{align}
Here we have used the abbreviated notation $X_i\defeq X_i(\xi)$ and $\hat{X}_i\defeq X_i(\hat{\xi})$ and $D\defeq \hat{X}_b X_b^{-1}-\hat{X}_a X_a^{-1}$. What we still assume is that $X_a(t)$ and $X_b(t)$ are invertible. However, this is merely a consequence of the one-to-one correspondence between initial data and solutions near any hypersurface of the foliation. We also assume that $D$ is invertible without being able to offer here an a priori justification.

\section{Using Green's Functions}
\label{sec:green}

Various operators appearing in our treatment can be made more explicit by using specific Green's functions of the Klein-Gordon equation.

The inverse of the symplectic structure (\ref{eq:sympl}) is a function $G:(I\times\R^3)\times (I\times\R^3)\to\R$ which is antisymmetric in its two arguments and satisfies the condition,
\be
 2\Omega(\phi,G_y)=\phi(y),\quad\text{where}\quad G_y(x)\defeq G(x,y),
\ee
and where $\phi$ is an arbitrary classical solution (see \cite{Woo:geomquant}). Indeed, this implies that $G$ is actually a Green's function and solves the Klein-Gordon equation in both of its arguments. Furthermore, it is easy to see that it satisfies the boundary conditions
\begin{gather}
 G((t,\underline{x}),(t,\underline{y}))=0\quad\forall \underline{x},\underline{y}\in\R^3,\forall t\in I,\\
 \sqrt{|g^{(3)}(t,\underline{x})g^{0 0}(t,\underline{x})|}\,
 \partial_\tau G((\tau,\underline{x}),(t,\underline{y}))|_{\tau=t}=-\delta^{(3)}(\underline{x}-\underline{y})\quad\forall \underline{x},\underline{y}\in\R^3,t\in I .
\end{gather}

Consider now the operators $X_a$ and $X_b$ of Section~\ref{sec:classical}. As a first step we express them via integral kernels $\tilde{X}_a$ and $\tilde{X}_b$, i.e.,
\be
 (X_i (t)\eta_i)(\underline{x})=\int\xd^3\underline{y}\,\tilde{X}_i(t,\underline{x},\underline{y}) \eta_i (\underline{y}) .
\ee
In order to relate these to a Green's function we have to specify what the ``initial'' data $\eta_a$, $\eta_b$ actually represent. We consider the following choice here: $\eta_a$ specifies the value of the field on the hypersurface $t=t_0$ while $\eta_b$ specifies the normal derivative of the field on the same hypersurface:
\begin{align}
\eta_a(\underline{x}) & =\phi(t_0,\underline{x}),\\
\eta_b(\underline{x}) & =\left.\sqrt{|g^{0 0}(t_0,\underline{x})|}\,\partial_\tau \phi(\tau,\underline{x})\right|_{\tau=t_0} .
\end{align}
 We obtain
\begin{align}
 \tilde{X}_a(t,\underline{x},\underline{y}) & = \left. -\sqrt{|g^{(3)}(t_0,\underline{y})g^{0 0}(t_0,\underline{y})|}\,\partial_\tau G((\tau,\underline{y}),(t,\underline{x}))\right|_{\tau=t_0},\\
 \tilde{X}_b(t,\underline{x},\underline{y}) & = \sqrt{|g^{(3)}(t_0,\underline{y})|}\,
 G((t_0,\underline{y}),(t,\underline{x})),
\end{align}
where $G$ is precisely the Green's function considered above.

The operators reconstructing a classical solution from its values on two hypersurfaces $t=\xi$ and $t=\hat{\xi}$ of the foliation as in equation (\ref{eq:solbdyval}) can also be related directly to a certain Green's function, different from $G$. (See \cite{Dop:tomschwing}, where such methods were used in a related context.) As a first step we write these operators in terms of integral kernels,
\begin{align}
\left(\frac{\Delta(t,\hat{\xi})}{\Delta(\xi,\hat{\xi})}\varphi\right)(\underline{x})
& =\int\xd^3\underline{y}\, K_{[\xi,\hat{\xi}]}(t,\underline{x},\underline{y}) \varphi(\underline{y}),\\
\left(\frac{\Delta(\xi,t)}{\Delta(\xi,\hat{\xi})}\hat{\varphi}\right)(\underline{x})
& =\int\xd^3\underline{y}\, L_{[\xi,\hat{\xi}]}(t,\underline{x},\underline{y}) \hat{\varphi}(\underline{y}) .
\end{align}
Now consider the symmetric Green's function $G_{[\xi,\hat{\xi}]}:(I\times\R^3)\times (I\times\R^3)\to\R$ with the following boundary conditions:
\begin{align}
 G_{[\xi,\hat{\xi}]}((\xi,\underline{x}),(t,\underline{y})) & =0\quad \forall \underline{x},\underline{y}\in\R^3,\forall t\in I,\\
 G_{[\xi,\hat{\xi}]}((\hat{\xi},\underline{x}),(t,\underline{y})) & =0\quad \forall \underline{x},\underline{y}\in\R^3,\forall t\in I.
\end{align}
The integral kernels $K$ and $L$ can then be written as
\begin{align}
 K_{[\xi,\hat{\xi}]}(t,\underline{x},\underline{y}) & = \left. \sqrt{|g^{(3)}(\xi,\underline{y})g^{0 0}(\xi,\underline{y})|}\,\partial_\tau G_{[\xi,\hat{\xi}]}((\tau,\underline{y}),(t,\underline{x}))\right|_{\tau=\xi},\\
 L_{[\xi,\hat{\xi}]}(t,\underline{x},\underline{y}) & = \left. -\sqrt{|g^{(3)}(\hat{\xi},\underline{y})g^{0 0}(\hat{\xi},\underline{y})|}\,\partial_\tau G_{[\xi,\hat{\xi}]}((\tau,\underline{y}),(t,\underline{x}))\right|_{\tau=\hat{\xi}}.
\end{align}
The operators $W_{[\xi,\hat{\xi}]}^{(i,j)}$ can also be written as integral kernels in a nicely symmetric way. Define
\be
 \left(W_{[\xi,\hat{\xi}]}^{(i,j)}\psi\right)(\underline{x})=\int\xd^3\underline{y}\, W_{[\xi,\hat{\xi}]}^{(i,j)}(\underline{x},\underline{y}) \psi(\underline{y}).
\ee
Then, combining equations (\ref{eq:W2}) with the expressions above yields,
\begin{align}
 W_{[\xi,\hat{\xi}]}^{(i,j)}(\underline{x},\underline{y})=\left.
-(-1)^{i-j}\sqrt{|g^{(3)}(\xi_i,\underline{x})g^{0 0}(\xi_i,\underline{x})|}
\sqrt{|g^{(3)}(\xi_j,\underline{y})g^{0 0}(\xi_j,\underline{y})|}
\,\partial_\tau \partial_\sigma G_{[\xi,\hat{\xi}]}((\tau,\underline{y}),(\sigma,\underline{x}))\right|_{\tau=\xi_j, \sigma=\xi_i} .
\end{align}
Here we have set $\xi_1\defeq\xi$ and $\xi_2\defeq \hat{\xi}$. The relative sign when $i\neq j$ originates from the opposite orientation of the two hypersurfaces as boundaries of the intermediate region.

\begin{acknowledgments}

This work was supported in part by CONACyT grant 49093.

\end{acknowledgments}

\bibliographystyle{amsordx}
\bibliography{stdrefs}

\end{document}